# Selective amplification of primary exciton in MoS$_2$ monolayer

Hyun Seok Lee[1,*], Min Su Kim[1], Youngjo Jin[1,2], Gang Hee Han[1], Young Hee Lee[1,2,3,†] and Jeongyong Kim[1,2,‡]

[1]*Center for Integrated Nanostructure Physics (CINAP), Institute for Basic Science (IBS), Sungkyunkwan University, Suwon 440-746, Korea.*

[2]*Department of Energy Science, Sungkyunkwan University, Suwon 440-746, Korea*

[3]*Department of Physics, Sungkyunkwan University, Suwon 440-746, Korea.*

E-mail: [*]hs.lee@skku.edu, [†]leeyoung@skku.edu, [‡]j.kim@skku.edu

## Abstract

Optoelectronics applications for transition-metal dichalcogenides are still limited by weak light absorption and their complex exciton modes are easily perturbed by varying excitation conditions, because they are inherent in atomically thin layers. Here, we propose a method of selectively amplifying the primary exciton (A$^0$) among the exciton complexes in monolayer MoS$_2$ via cyclic re-excitation of cavity-free exciton-coupled plasmon propagation. This was implemented by partially overlapping a Ag nanowire (NW) on a MoS$_2$ monolayer separated by a thin SiO$_2$ spacer. Exciton-coupled plasmons in the NW enhance the A$^0$ radiation in MoS$_2$. The cumulative amplification of emission enhancement by cyclic plasmon travelling reaches ~20-fold selectively for the A$^0$, while excluding other B exciton and multiexciton by significantly reduced band-filling, without oscillatory spectra implying plasmonic cavity effects.

Two advantages of two-dimensional (2D) transition-metal dichalcogenide (TMD) semiconductors in 2D optoelectronics are band-gap tuning via controlling atomic layers and hybridizations, and on-chip integration via direct growth [1-4]. Manipulating exciton emissions by controlling light-emitter interactions is key to exciton engineering [5]. However, their inherent drawback is weak light absorption by their atomically thin layer [6]. The use of the local-field enhancement effect of localized surface plasmons via hybridization with metal nanostructures is a promising way to enhance the optoelectronic performance of TMDs [1,6-8]. However, this approach is challenging because coherent tuning between the plasmon resonance and the optical wavelength must be precisely engineered [7-10], and the mechanism of the interaction between surface plasmon polaritons (SPPs) and the intricate excitons of TMDs must be understood [11-17].

Here, we propose a cavity-free method to enhance the exciton emission performance of TMD semiconductors under varying excitation laser power ($P_{ex}$) without sacrificing peak quality and shape. We used nanowire (NW)-TMD emitter hybrids in this study (Methods [18]). The Ag-NW partly overlapping on the monolayer $MoS_2$ from which it was separated by a $SiO_2$ (10 nm) spacer to prevent band-pinning, doping, and photoluminescence (PL) quenching from direct metal-semiconductor contact [19,20]. Figure 1(a) shows a schematic and cross-sectional view of the experimental setup. $A^0$ (primary neutral A exciton, ~1.88 eV), A' (multiexciton, ~1.84 eV), and B (B exciton, ~2.02 eV) indicate typical exciton complexes of $MoS_2$ monolayers [11,13-15]. "On-NW" is the NW-$MoS_2$ overlapping region (NMOR), and "Off-NW" is the bare $MoS_2$ region.

The PL image in Fig. 1(b) exhibits strong red emission at the laser input position (LIP) indicated by a green arrow and weak but still prominent emission at the NW-end position (NWEP), implying that $MoS_2$ excitons are coupled to SPPs and propagate along the NW

[20-22]. The PL signals were collected at the same LIP indicated by a white arrow. Figure 1(c) shows the normalized PL spectra that were deconvoluted using a Lorentzian function. The unknown X peak for Off-NW is presumably a localized state due to defects or impurities [23]. At all five $P_{ex}$ levels (5-500 µW), only the $A^0$ observation for On-NW is markedly different from the PL spectra for Off-NW which have three exciton modes. For On-NW, the A-peak position of ~1.88 eV and full-width at half-maximum (FWHM) of ~50 meV remain unchanged and independent of $P_{ex}$. However, for Off-NW, as $P_{ex}$ increases, the A-peak center position redshifts considerably because the A' dominates the A-peak, and the intensities of A' and B increase.

Figure 1(d) compares the PL spectra for On-NW and Off-NW at $P_{ex}$ = 100 µW. Notably, the A-peak intensity ($I_A$) for On-NW consisting of only the $A^0$ is dramatically enhanced compared with that for Off-NW consisting of the $A^0$ and A' [Fig. 1(c)]. The A-peak enhancement factor is defined as $\varepsilon = I_{PL}^{On}/I_{PL}^{Off}$, where $I_{PL}^{On}$ and $I_{PL}^{Off}$ are the maximum $I_A$ for On-NW and Off-NW, respectively, and the $\varepsilon$-factor can reach ~20. Figure 1(e) shows the log-log scale intensity-power curves for A and B peaks in Off-NW. The integrated PL intensity ($I_{PL}$) is approximately equal to $(P_{ex})^m$, where $m$ denotes exponent. For the $I_A$, $m$ ~ 0.9 at $P_{ex}$ = 5-100 nW, where only the $A^0$ is identified via a single Lorentzian fit (SLF) [Fig. 1(f)]. However, at $P_{ex}$ > 100 nW, the $I_A$ is saturated and $m$ is degraded to ~0.6 as the A' starts to evolve and dominate the A-peak [Fig. 1(c), Off-NW]. As the A-peak starts being saturated over 100 nW, the B-peak intensity ($I_B$) also starts to emerge and increase in response to $P_{ex}$ with $m$ ~ 1.3. Therefore, for Off-NW, the exclusive emergence of the $A^0$ at extremely low $P_{ex}$ [Fig. 1(f)] strongly contrasts with the formation of exciton complexes (A' and B) at high $P_{ex}$ [Fig. 1(c)]; this is attributed to the result of the band-filling effect for excitons [24-27]. Details are given in Note 1 [18]. Moreover, because the FWHM

(~50 meV) for Off-NW at low $P_{ex}$ [Fig. 1(f)] is the same as that for On-NW [Fig. 1(c)], the origin of PL enhancement (PLE) for On-NW [Fig. 1(d)] cannot be associated with the cavity resonance effect revealing oscillatory fringes [28].

To determine the role of NWs in $A^0$ enhancement, the NW length effect was investigated in fully and partially overlapping NW samples, where the PL signals were collected at the same LIP. $x_M$ is the length of the fully overlapping NW (FONW), while for the partially overlapping NW (PONW), $x$ is the effective length of bare NW measured from the NWEP to the LIP [Fig. 2(a)]. The $\varepsilon$-factors were measured for numerous devices [Fig. 2(b)]. For FONW samples, the $\varepsilon$-factor (mean ~1.3) were appreciably negligible for various $x_M$ values (Section 1 [18]). However, for the PONW samples, the considerable enhancement (~20 times for $x$ ~ 3 μm) decayed exponentially (decay length ~ 6 μm) as a function of $x$ despite the significant fluctuation in $\varepsilon$ values, which is attributed to the variable sample conditions, i.e., NW quality, laser focusing, and SPP reflectivity at the NWEP (Section 4 [18]). The lack of $x_M$-dependence of $\varepsilon$-factor in the FONW samples implies that the selective $A^0$ enhancement is associated with longitudinal-mode SPP (L-SPP) propagation along the $x$.

To characterize the propagation behavior of exciton-coupled (EC)-SPPs in NWs, PL signals for longer NWs ($x$ ~ 6.5 μm) were collected at the LIP [Fig. 3(b)] and at NWEP [Fig. 3(c)] during laser illumination at the NMOR. In Fig 3(a), the left image indicates the LIP (green arrow) and the right image indicates the PL collecting positions for the LIP and the NWEP (white arrows). At the LIP, $\varepsilon$-factor reaches ~7 selectively for the $A^0$ [Fig. 3(b)]. The $A^0$ selectivity is confirmed by subtracting the PL spectrum for Off-NW from that of On-NW (which yields ΔPL) and the $A^0$ is identified exclusively via a SLF. [Fig. 3(b), inset].

In linear-scale PL spectra collected at the NWEP [Fig. 3(c), inset], the $A^0$ also dominates A-peak for all $P_{ex}$. Interestingly, the log-scale plot of the spectra [Fig. 3(c)] shows oscillatory fringes at photon energy lower than that of $A^0$ ($\hbar\omega < \hbar\omega_{A0}$) and no fringes at $\hbar\omega \geq \hbar\omega_{A0}$. The SPPs with $\hbar\omega < \hbar\omega_{A0}$ (involving A′) coupled from the A-peak tail form Fabry-Pérot cavity modes of poor quality (Section 2 [18]) in which excitons travel through NWs as standing waves that generally accompany oscillatory spectra [9,20]. This cavity signal is observed only at the NWEP, because the scattering of L-SPPs is negligible at the LIP (midsection of NW) with the symmetrical NW geometry in the axial direction (Section 3 [18]). Conversely, at the NMOR, the SPPs with $\hbar\omega \geq \hbar\omega_{A0}$ are reabsorbed into the $MoS_2$ layer after their round trip and then lose their energies by PL emission in $MoS_2$ (Section 4 and 5 [18]); thus, the cavity modes are not constructed. Although the SPPs with $\hbar\omega < \hbar\omega_{A0}$ exhibit some cavity effect, they do not contribute to $A^0$ amplification because they are not absorbed by the $MoS_2$ layer. Therefore, this selective $A^0$ amplification phenomenon at the LIP differs from the PLE via the plasmonic antenna effect [1,8] (Section 4 [18]) or the conventional cavity effect [9,20].

Remarkably, the PONWs play a key role in selective $A^0$ enhancement. Furthermore, this enhancement becomes more prominent as $x$ decreases. In Fig. 4(a), the input laser generates excitons in the $MoS_2$ and the excitons couple with SPPs in the NW. The EC-SPPs propagate through the NW and return to their LIP after plasmon ohmic loss in the NW, which is the result of the imaginary part of the dielectric function of Ag and loss due to scattering at the NWEP [29]. The SPPs returned to their LIP enhance the spontaneous emission (SE) rate of the generated $MoS_2$ excitons, resulting in a gain in exciton emission ($\gamma > 1$) [30]. The SE enhancement via plasmonic resonant coupling with a metal nanostructure is known as the Purcell enhancement (PE) effect [10]. Conversely, SPPs with a tightly confined mode enhance SE due to the nonresonant PE effect, even without the longitudinal cavity effect [10,30]. The nonresonant PE factor is given by $F_{NP} = \Gamma_{SPP}/\Gamma_X \propto (\lambda_e/d)^3$, where $\Gamma_{SPP}$ and $\Gamma_X$ are the radiative decay rate for On-NW and Off-NW, respectively; $\lambda_e$ is the exciton wavelength in free space, and $d$ is the thickness of spacer [21,30]; and thus, $\gamma \propto F_{NP}$. Because $\Gamma_{SPP} > \Gamma_X$, the enhanced SE can restrict the band-filling effect (Note 1 and 2 [18]). Therefore, mostly the $A^0$ is generated for On-NW, as discussed in Fig. 1.

The enhanced $A^0$ emission due to the returned SPPs recouples to the SPPs at NMOR, resulting in SPP intensity enhancement (SPP enhancement factor, $\gamma' > 1$) [9,31]. This agrees with the $A^0$ dominance in the EC-SPPs monitored at the NWEP [Fig. 3(c)]. Therefore, the accumulation of selective $A^0$ re-excitation via cyclic round trips of the SPPs significantly amplifies the $A^0$ intensity at their LIP through the repeated PL gain process [Fig. 4(b) and (c)]. Details of the analytical model are given in Note 2 [18]. The EC-SPPs from the MoS$_2$ layer via laser excitation, defined as $i_0$, propagate along the left ($x$) and the right ($x'$: NMOR) regions of the NW and undergo SPP ohmic loss, which is proportional to $0.5 i_0 e^{(-x/L)}$ and $0.5 i_0 e^{(-x'/L')}$, respectively, where $L$ and $L'$ are the decay lengths of the SPPs in the $x$ and $x'$ directions, respectively, and the factor of 0.5 is applied because it is assumed that the equivalent amount of SPPs travel in each NW direction [Fig. 4(a)]. In this situation, the SPPs that are coupled directly with the input laser at the NW midsection are ignored, because L-SPPs are not activated as they are at the NWEP [32] (Sections 3 and 4 [18]). The EC-SPPs are reflected at the NWEP with a reflectivity of $r \sim 0.25$ and returned to the LIP [29].

The intensity of the returned SPPs after one round trip is given by $i_1^+ = 0.5 r i_0 e^{(-2x/L)} + 0.5 r i_0 e^{(-2x'/L')}$. We assumed $L \gg L'$ due to the additional loss of SPPs resulting from their reabsorption and reemission in the MoS$_2$ layer during propagation along the $x'$ direction (Section 5 [18]). Therefore, $i_1^+ \sim 0.5 r i_0 e^{(-2x/L)} = i_0 g$. Figure 4(c) shows the amplification flow via repeated interaction between excitons and SPPs as the number of SPP travelling cycles increases. The $i_1^+$ re-excites additional exciton ($\delta_1^{ex}$) in MoS$_2$ and $\delta_1^{ex} = \gamma i_1^+$. $\delta_1^{ex} = \delta_1^{PL} + \delta_1^{SPP}$, where $\delta_1^{PL} = (1-\rho)\delta_1^{ex}$ is the enhanced PL and $\delta_1^{SPP} = \rho \delta_1^{ex} = \rho \gamma'(\gamma i_1^+)$ is recoupled to SPPs with enhanced intensity ($\gamma'$) for the second round of travel, where $\rho < 1$ denotes the partial ratio of $\delta_1^{ex}$. After $n$ travel cycles, the total SPP intensity enhancement and the total PLE are given by $\Delta i^+ = i_0 g \sum_{n=0}^{\infty} (\gamma' \gamma \rho g)^n$ and $\Delta \delta^{PL} = (1-\rho) \gamma i_0 g \sum_{n=0}^{\infty} (\gamma' \gamma \rho g)^n$, respectively. From the simulation result, $L \sim 12$ μm for the A$^0$ (Section 3 [18]). When $x > 3$ μm, $g < \sim 0.08 \ll 1$. For convergence, $\gamma' \gamma \rho g < 1$. By Taylor expansion, $\sum_{n=0}^{\infty} (\gamma' \gamma \rho g)^n \sim 1/(1-\gamma' \gamma \rho g)$, so $\Delta \delta^{PL} \sim (1-\rho) \gamma i_0 g / (1-\gamma' \gamma \rho g)$ and $\Delta i^+ \sim i_0 g / (1-\gamma' \gamma \rho g)$. For the first SPP propagation, the SPP intensity is given by $i_1^s = (1-r) 0.5 i_0 e^{(-x/L)} = i_0 g'$. After $n$ travel cycles, the total scattered SPPs at the NWEP is given by $\Delta i^s = i_0 g' \sum_{n=0}^{\infty} (\gamma' \gamma \rho g)^n \sim i_0 g' / (1-\gamma' \gamma \rho g)$. Because the ΔPL intensity at the LIP is stronger than that of the scattered SPPs at the NWEP by more than 20-fold [Fig. 3], $\Delta \delta^{PL} / \Delta i^s \sim (1-\rho) \gamma i_0 g / (i_0 g') \sim 20$, which yields $\gamma \sim C/(1-\rho)$ with $C \sim 100$. Eventually, $\Delta \delta^{PL} \sim C i_0 g / (1-C \beta g)$, where $\beta = \gamma' \rho / (1-\rho)$. The exciton-SPP conversion efficiency was determined experimentally as $\eta = i_0/(I_{PL}^{Off}+i_0) \sim 0.32$ [22], and thus $i_0 = \eta/(1-\eta)I_{PL}^{Off} \sim 0.5 I_{PL}^{Off}$. The $\varepsilon$-factor is approximated as

$$\varepsilon = \frac{I_{On}^{PL}}{I_{Off}^{PL}} \sim \frac{I_{Off}^{PL} + \Delta\delta^{rad}}{I_{Off}^{PL}} \sim 1 + \frac{Ci_0 g}{1 - C\beta g I_{Off}^{PL}} \sim 1 + \frac{6.4 e^{-x/6}}{1 - 12.9\beta e^{-x/6}}. \quad (1)$$

Figure 4(d) presents the experimental data for the PONW samples from Fig. 2(b) and plots of Eq. (1) in response to various $\beta$. With $\beta$ in the range of $0.001 \leq \beta \leq 0.1$, the model agrees well with the data. For $\beta \sim 0.1$, $\rho \sim 0.1$ to satisfy $\gamma' > 1$ and the maximum $\varepsilon$-factor becomes ~20. For $\beta \sim 0.001$, $\rho \sim 0$, $\gamma \sim C \sim 100$, and the PLE effect is very weak; thus, the $\varepsilon$-factor goes to <5 because the term with $\beta$ in the denominator of Eq. (1), which implies the cyclic $A^0$ accumulation effect, is negligible. When $\rho$ is not negligible but small, therefore, the PLE effect is prominent due to the prominent contribution of the cyclic $A^0$ accumulation effect.

Our method for selectively amplifying the primary exciton with cavity-free tunability can open a shortcut to realize high-performance TMD optoelectronics. Moreover, our finding that exciton-coupled plasmons excite mostly the primary exciton among exciton complexes provides a deeper understanding of the complicated emission behavior of excitons in different TMDs.

Acknowledgement: This work was supported by IBS-R011-D1.

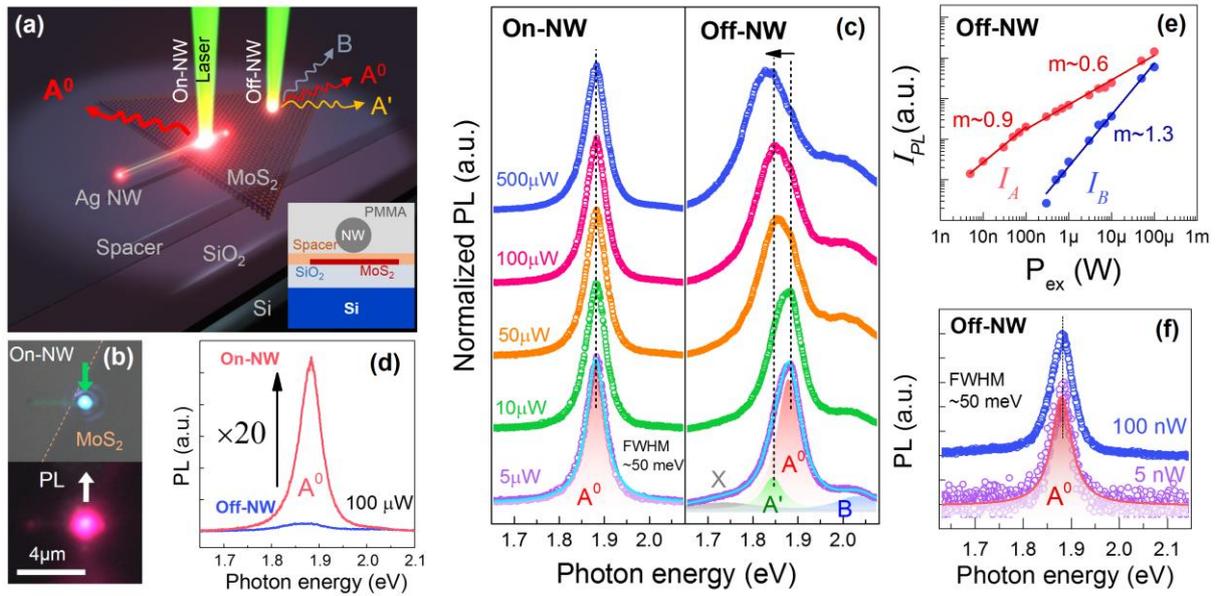

FIG 1. (a) Schematic of the experimental setup with a side view of the hybrid. PL signals were collected from the NMOR (On-NW) and from the bare $MoS_2$ (Off-NW) that were excited by an input laser. (b) (top) Optical micrograph showing the LIP (green arrow) and (bottom) PL image showing the collection position (white arrow) of the PL signal at the same LIP. NW length: ~4 µm. Effective NW length from the NWEP to the LIP: ~3 µm. (c) Normalized PL signals as a function of $P_{ex}$ for On-NW and Off-NW, with examples of Lorentzian deconvolution at $P_{ex} = 5$ µW. (d) PL spectra for On-NW and Off-NW at $P_{ex} = 100$ µW. (e) The log-log scale PL intensity ($I_{PL}$) as a function of $P_{ex}$ derived from the PL spectra for Off-NW. (f) The PL spectra for Off-NW at $P_{ex} = 5$ nW and 100 nW and SLF for $P_{ex} = 5$ nW identified as $A^0$.

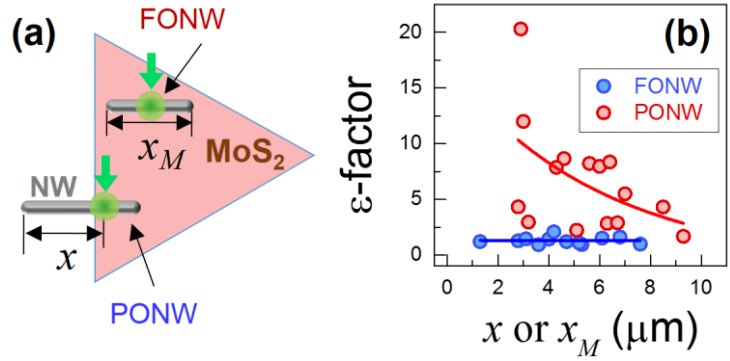

FIG. 2. (a) Schematics for two different sample configurations: fully overlapping NW (FONW) and partially overlapping NW (PONW) on MoS$_2$ flakes. $x_M$: the length of the FONW. $x$: the effective length for the PONW from the end of the bare NW to the LIP (green arrow). (b) $\varepsilon$-factor (maximum $I_A$ for On-NW divided by that of Off-NW) as a function of $x$ or $x_M$. At $P_{ex}$ = 100 μW, the PL signals were collected at the same LIP for numerous devices with the two different sample configurations.

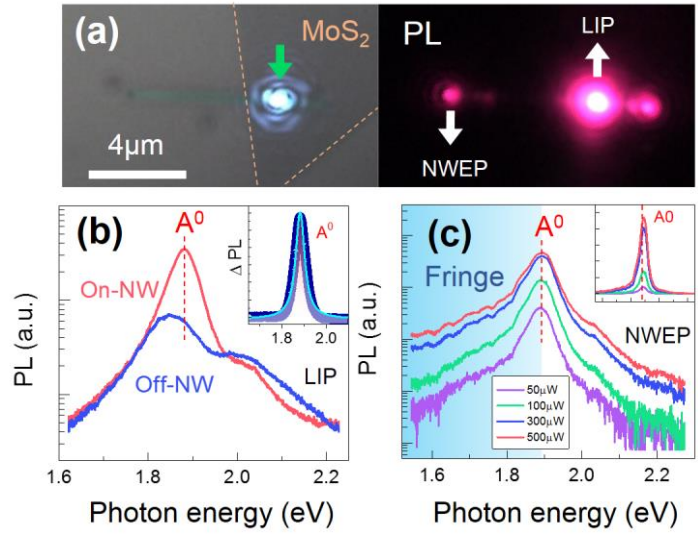

FIG. 3 (a) Optical micrograph (left) illustrating MoS$_2$ flake position and PL image (right). $x$: ~6.5 μm; NW length: ~8 μm; green arrow: LIP; white arrow: PL signal collection position. (b) PL spectra collected at the LIP. Log-scale PL spectra comparison between On-NW and Off-NW with ε-factor of ~7 at $P_{ex}$ = 500 μW. Inset: PL for Off-NW subtracted from that of On-NW and its SLF identified as A$^0$ implying only A$^0$ enhancement for On-NW. (c) PL spectra collected at the NWEP as a function of $P_{ex}$. Log-scale PL spectra reveal oscillatory fringes that are apparent at $\hbar\omega < \hbar\omega_{A0}$. Inset: linear scale.

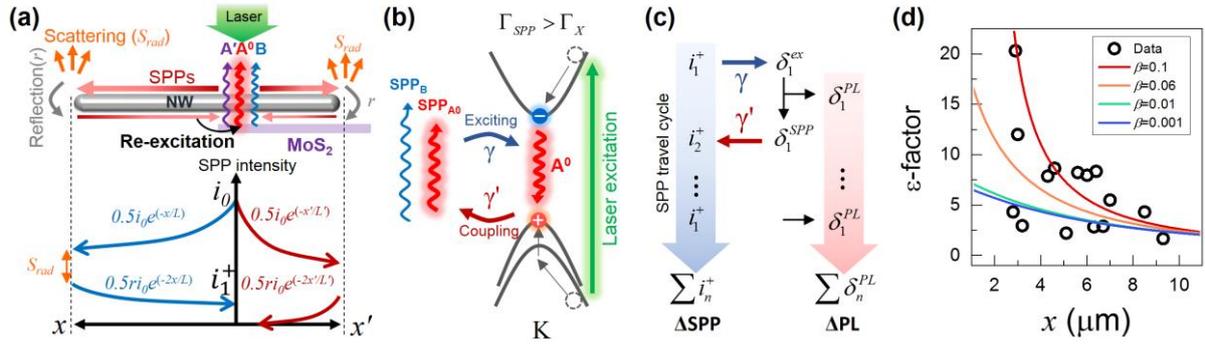

FIG. 4 (a) Complex excitons ($A^0$, $A'$, and B) excited under an intense laser coupled to the SPPs at the LIP. The travelling SPPs lose their intensity via ohmic loss along the NW and scattering loss ($S_{rad}$) at the NWEP. Reflected SPPs selectively re-excite only the $A^0$. (b) SPP cyclic travelling enhances the $A^0$ emission by restricting the band-filling effect in $MoS_2$ excitons. (c) PL and SPP amplification flow via repeated interaction between excitons and SPPs as the travelling cycles increase. (d) Comparison of the experimental data for the PONW samples in Fig. 2(b) with the plots of Eq. (1) for various $\beta$ as a function of $x$.

# Supplemental Materials

## Selective amplification of primary exciton in monolayer MoS$_2$

Hyun Seok Lee[1,*], Min Su Kim[1], Youngjo Jin[1,2], Gang Hee Han[1], Young Hee Lee[1,2,3,†] and Jeongyong Kim[1,2,‡]

[1]*Center for Integrated Nanostructure Physics (CINAP), Institute for Basic Science (IBS), Sungkyunkwan University, Suwon 440-746, Korea*
[2]*Department of Energy Science, Sungkyunkwan University, Suwon 440-746, Korea*
[3]*Department of Physics, Sungkyunkwan University, Suwon 440-746, Korea*

Corresponding author: [*]hs.lee@skku.edu, [†]leeyoung@skku.edu, [‡]j.kim@skku.edu

## Methods

Monolayer $MoS_2$ flakes were synthesized on a $SiO_2$ (300 nm) / Si substrate using a vapor phase reaction [33]. Poly (methylmethacrylate) (950K PMMA, MicroChem Corp., 4% in chlorobenzene) was spin-coated onto $MoS_2$-grown samples and dried under ambient conditions. The samples were soaked in a 1 M KOH solution for several minutes to etch the $SiO_2$ surface. Subsequently, the detached PMMA/$MoS_2$ layers were washed with de-ionized water and transferred to $SiO_2$ (300 nm)/Si wafers. Finally, the PMMA layer was removed using acetone. To prevent band-pinning, doping, and photoluminescence (PL) quenching due to direct metal/semiconductor contact [19,20,34,35], a $SiO_2$ (10 nm) spacer layer was deposited onto the samples via an electron beam evaporation method. The diameters (~110 nm) of the silver (Ag) NWs (Sigma-Aldrich Corp., dispersed in isopropyl alcohol) were confirmed by atomic force microscopy after sonication to shorten wire lengths and spread onto the samples. Finally, a thick layer (~400 nm) of PMMA was utilized to cover the samples to prevent NW degradation from the atmosphere.

For PL experiments, the lab-constructed confocal microscope was used. A laser beam (wavelength, $\lambda$ = 514 nm) was focused on the sample by an objective lens (x100, numerical aperture, 0.9) for PL experiments. PL spectra from the samples were collected in a pinhole detector (positioned by a micromanipulator). The spectra were recorded using a spectrometer and a cooled charge-coupled device camera.

## Note 1: Band-filling effect in excitons

For Off-NW, the exclusive emergence of the $A^0$ at extremely low $P_{ex}$ [Fig. 1(f)] strongly contrasts with the formation of complex excitons (A' and B) at high $P_{ex}$ [Fig. 1(c)]; this is attributed to the result of the band-filling (or phase-space filling) effect for excitons [24-27]. In semiconductors, hot carriers excited by laser illumination are thermalized to the lowest exciton energy level and recombine via radiative or nonradiative decay. Under intense light, when the optical carrier generation rate is much lower than the recombination rate, the thermalized carriers accumulate from the band edge and the quasi-Fermi level shifts deeper into the conduction and valance bands in a steady state [24]. This band-filling effect also applies to excitons [25-27]. In the previous report [15], the steady-state solutions of the rate equation for the population of $A^0$ ($N_X$) and A' ($N_M$) for MoS$_2$ were $N_X = G/(\Gamma_X + \theta)$ and $N_M = \theta G/(\Gamma_X \Gamma_M + \theta \Gamma_M)$, where $G$, $\theta$, $\Gamma_X$, and $\Gamma_M$ are the exciton generation rate, the A' formation rate and the radiative decay rates for $A^0$ and A', respectively. Because experimental results have shown that $\theta \gg \Gamma_X$ [15], the equations become $N_X \sim G/\theta$ and $N_M \sim G/\Gamma_M$. For $P_{ex} > 5$ μW in Off-NW of Fig. 1(c), $N_X$ decreases and $N_M$ increases in response to $P_{ex}$; therefore, $\theta \propto P_{ex}$ and $G \propto P_{ex}$ is satisfied. For $P_{ex} \leq 100$ nW, $\theta$ is negligible because there are not enough excitons to fill the exciton states; thus, $N_X \sim G/\Gamma_X$ and $N_M \sim 0$, as shown in Fig. 1(f). Conversely, for $P_{ex} > 100$ nW, as $P_{ex}$ increases, $\theta$ also increases because the enhanced probability of charge-charge interactions in electron-hole plasma in the occupied exciton states may create more A', e.g., trions or bi-excitons [14-16,36] as shown in Fig. 1(c) for Off-NW and Fig. 1(e). The probability of forming charged excitons such as trions because of the intrinsic n-doped features of MoS$_2$ samples cannot be ignored [14,15]. The B peak does not emerge at low laser power but prominently appears at high laser power, because the B state can be filled after the A state has been sufficiently filled. Moreover, the exclusive emergence of A' and B excitons at only high laser power and the $A^0$ exciton from the band-filling effect contrast with absorption spectroscopy results in which both A and B excitons always appeared [13,27].

**Note 2: Details of the analytical model for cyclic amplification**

The exciton-coupled SPPs from the MoS$_2$ layer, derived via laser excitation and defined as $i_0$, propagate along the left (*x*: bare NW region) and the right (*x'*: NW/MoS$_2$ overlapping region) sides of the NW. During propagation, the SPPs undergo ohmic loss, which is proportional to $0.5i_0 e^{(-x/L)}$ and $0.5i_0 e^{(-x'/L')}$, respectively; *L* and *L'* are the decay lengths of the SPPs for the *x* and *x'* sections of the NW and the factor of 0.5 is applied under the assumption that the equivalent amount of SPPs travel in each direction [Fig. 4(a)]. In Fig. 4(a), the SPPs that coupled directly with the input laser in the NW midsection are ignored because the longitudinal SPP modes are not activated as they are in the NW end [32]. This was reconfirmed by simulations and experiments (Sections 3 and 4). The exciton-coupled SPPs are reflected with a reflectivity *r* of ~0.25 at the NW ends and returned to the laser input position [29]. After one round trip, the intensity of the returned SPPs is given by $i_1^+ = 0.5 r i_0 e^{(-2x/L)} + 0.5 r i_0 e^{(-2x'/L')}$. Here, we assumed that $L \gg L'$ and the amplification effect for the *x'* direction is negligible because of the additional loss of SPPs due to their reabsorption and re-emission in the MoS$_2$ layer during propagation (Section 5). Additionally, this was shown by the negligible PL enhancement for the fully overlapping samples found by our experiments [Fig. 2(a)]; therefore, $i_1^+ \sim 0.5 r i_0 e^{(-2x/L)} = i_0 g$.

In a longitudinal Fabry-Pérot cavity for cylindrical plasmonic waveguides, the Purcell enhancement factor is given by $F_P = \Gamma_{cavity} / \Gamma_0 \propto Q(\lambda_0^3 / V_{eff})$, where $\Gamma_{cavity}$ and $\Gamma_0$ are the radiative decay rate of the emitter with and without a cavity, $\lambda_0$ is the free-space wavelength, and *Q* is the quality factor of the cavity [10]. In our case, a nonresonant Purcell enhancement factor without the cavity effect is given by $F_{NP} = \Gamma_{SPP} / \Gamma_X \propto (\lambda_e / d)^3$, where $\Gamma_{SPP}$ and $\Gamma_X$ are the radiative decay rate for On-NW and Off-NW in the NW/MoS$_2$ hybrid, respectively; $\lambda_e$ is the exciton wavelength in free space, and *d* is the thickness of the spacer between the NW and the emitter [21,30], and thus, $\gamma \propto F_{NP}$. Because $\Gamma_{SPP} > \Gamma_X$, the enhanced spontaneous emission can restrict the band-filling effect and thus $\theta$ decreases. When $\theta \ll \Gamma_{SPP}$ and $\theta \sim 0$, then $N_X \sim G / \Gamma_{SPP}$ and $N_M \sim 0$. Therefore, mostly the A$^0$ is generated on the NW/MoS$_2$ overlapping region (On-NW), as discussed in Fig. 1. Notably, among the excitons coupled to SPPs, the A'-coupled SPPs cannot re-excite the A$^0$ because their energy is lower than that of the $A^0$. As discussed earlier, the cavity mode of A' is generated due to the absence of A' reabsorption in the $MoS_2$ layer [Fig. 3(c) and Section 2].

Here, we consider only the fundamental waveguide mode because excitation of higher-order modes is difficult [37]. Note that the spontaneous emission enhancement is due to the high confinement of optical mode near the NW, not to the longitudinal cavity effect [21,30]. Therefore, exciton emission gain due to SPPs ($\gamma$) is proportional to $F_{NP}$, while the contribution of emission enhancement due to the longitudinal round trips of the SPPs is not involved in this effect. After the cyclic travelling of the SPPs, the enhanced exciton emission recouples with the SPPs and the SPP intensity is enhanced by $\gamma'$. Cyclic interactions of the enhanced SPP intensity with exctions eventually behave like the Q factor for the resonant cavity. Therefore, only $A^0$ enhancement of SPPs is observed at the NW end, as shown in Fig. 3(c).

Figure 4(c) shows the amplification flow via repeated interaction between excitons and SPPs as the number of SPP travelling cycles increases. The $i_1^+$ re-excites additional exciton ($\delta_1^{ex}$) in the $MoS_2$ and $\delta_1^{ex} = \gamma i_1^+$, which is then defined as $\delta_1^{ex} = \delta_1^{PL} + \delta_1^{SPP}$. While $\delta_1^{PL} = (1-\rho)\delta_1^{ex}$ is the enhanced PL measured in the laser input position, $\delta_1^{SPP} = \rho\delta_1^{ex}$ is recoupled to the SPPs with intensity enhancement $\gamma'$ for the second round trip, and $\delta_1^{SPP} = \rho\gamma'(\gamma i_1^+)$, where $\rho < 1$ denotes the partial ratio of $\delta_1^{ex}$. After the $n^{th}$ cycle, the SPP intensity enhancement is given by

$$i_1^+ = i_0 g,$$
$$i_2^+ = \rho\gamma'[\gamma i_1^+]g = \rho\gamma'[\gamma(i_0 g)]g = i_0 g(\gamma'\gamma\rho g),$$
$$i_3^+ = \rho\gamma'[\gamma i_2^+]g = \rho\gamma'[\gamma(i_0 g(\gamma'\gamma\rho g))]g = i_0 g(\gamma'\gamma\rho g)^2,$$
$$\vdots$$

$$i_n^+ = i_0 g(\gamma'\gamma\rho g)^{n-1}, \tag{S1}$$

and the total SPP intensity enhancement is given by

$$\Delta i^+ = \sum i_n^+ = i_0 g \sum_{n=0}^{\infty} (\gamma'\gamma\rho g)^n. \tag{S2}$$

After the $n^{th}$ cycle, the PL enhancement is given by

$$\delta_1^{PL} = (1-\rho)\delta_1^{ex} = (1-\rho)\gamma[i_1^+] = (1-\rho)\gamma[i_0 g],$$
$$\delta_2^{PL} = (1-\rho)\delta_2^{ex} = (1-\rho)\gamma[i_2^+] = (1-\rho)\gamma[i_0 g(\gamma'\gamma\rho g)],$$
$$\delta_3^{PL} = (1-\rho)\delta_3^{ex} = (1-\rho)\gamma[i_3^+] = (1-\rho)\gamma[i_0 g(\gamma'\gamma\rho g)^2],$$
$$\vdots$$
$$\delta_n^{PL} = (1-\rho)\gamma i_0 g(\gamma'\gamma\rho g)^{n-1}, \tag{S3}$$

and the total PL enhancement is given by

$$\Delta\delta^{PL} = \sum \delta_n^{PL} = (1-\rho)\gamma i_0 g \sum_{n=0}^{\infty}(\gamma'\gamma\rho g)^n. \tag{S4}$$

For the $A^0$ mode, $L \sim 12$ μm (Section 3). For a long NW ($x > 3$ μm), $g < \sim 0.08 \ll 1$. In our model, $\rho < 1$, $\gamma > 1$, and $\gamma' > 1$. For convergence, $\gamma'\gamma\rho g < 1$. By Taylor expansion, $\sum_{n=0}^{\infty}(\gamma'\gamma\rho g)^n \sim 1/(1-\gamma'\gamma\rho g)$, so

$$\Delta\delta^{PL} \sim \frac{(1-\rho)\gamma i_0 g}{1-\gamma'\gamma\rho g} \tag{S5}$$

and

$$\Delta i^+ \sim \frac{i_0 g}{1-\gamma'\gamma\rho g}. \tag{S6}$$

To estimate $\gamma$, we consider the intensity of the scattered SPPs at the NW end. For the first propagation of SPPs, $i_1^s = (1-r)0.5 i_0 e^{(-x/L)} = i_0 g'$. After the $n^{th}$ round trip,

$$i_1^s = i_0 g',$$
$$i_2^s = \rho\gamma'[\gamma i_1^+]g' = \rho\gamma'[\gamma(i_0 g)]g' = i_0 g'(\gamma'\gamma\rho g),$$
$$i_3^s = \rho\gamma'[\gamma i_2^+]g' = \rho\gamma'[\gamma(i_0 g(\gamma'\gamma\rho g))]g' = i_0 g'(\gamma'\gamma\rho g)^2,$$
$$\vdots$$
$$i_n^s = i_0 g'(\gamma'\gamma\rho g)^{n-1}, \tag{S7}$$

and the total scattered SPPs at the NW end is given by

$$\Delta i^s = \sum i_n^s = i_0 g' \sum_{n=0}^{\infty}(\gamma'\gamma\rho g)^n \sim \frac{i_0 g'}{1-\gamma'\gamma\rho g}. \tag{S8}$$

Because the ΔPL intensity at the input laser position is more than 20 times stronger than the intensity of the scattered PL at the NW end, when $r \sim 0.25$ and $x \sim 6.5$ μm [Fig. 3], Eq. (S5) divided by Eq. (S8) gives

$$\frac{\Delta\delta^{PL}}{\Delta i^s} \sim \frac{(1-\rho)\gamma i_0 g}{i_0 g'} \sim \frac{0.5(1-\rho)\gamma r e^{-2x/L}}{0.5(1-r)e^{-x/L}},$$

and with $L \sim 12$ μm as defined earlier from the simulation result,

$$\frac{\Delta\delta^{PL}}{\Delta i^s} \sim \frac{0.5 \times 0.25(1-\rho)\gamma e^{-x/L}}{0.5 \times (1-0.25)} \sim \frac{0.25(1-\rho)\gamma e^{-6.5/12}}{0.75} \sim 20. \quad (S9)$$

Then we can obtain $\gamma$,

$$\gamma \sim \frac{20 \times 0.75}{0.25 \times 0.58(1-\rho)} \sim C/(1-\rho), \quad (S10)$$

where $C \sim 100$. By inserting Eq. (S10) into Eq. (S5), we get

$$\Delta\delta^{PL} \sim \frac{(1-\rho)\dfrac{C}{(1-\rho)}i_0 g}{1-\dfrac{C\gamma'\rho}{(1-\rho)}g} \sim \frac{Ci_0 g}{1-C\beta g}, \quad (S11)$$

where $\beta = \gamma'\rho/(1-\rho)$ and $C\beta g < 1$. The exciton-SPP conversion efficiency was determined experimentally as $\eta = i_0/(I_{PL}^{Off}+i_0) \sim 0.32$ [22], so $i_0 = \eta/(1-\eta)I_{PL}^{Off} \sim 0.5 I_{PL}^{Off}$.

The $\varepsilon$-factor is approximated as

$$\varepsilon = \frac{I_{On}^{PL}}{I_{Off}^{PL}} \sim \frac{I_{Off}^{PL}+\Delta\delta^{rad}}{I_{Off}^{PL}} \sim 1+\frac{Ci_0 g}{1-C\beta g I_{Off}^{PL}} \sim 1+\frac{100 \times 0.5 g}{1-100\beta g} \sim 1+\frac{100 \times 0.5 \times (0.5 \times 0.25 e^{-2x/12})}{1-100\beta(0.5 \times 0.25 e^{-2x/12})}$$

and

$$\varepsilon \sim 1+\frac{6.4 e^{-x/6}}{1-12.9\beta e^{-x/6}}. \quad (S12)$$

Figure 4(d) shows the experimental data for the partially overlapping samples from Fig. 2(b) and plots of Eq. (S12) for different $\beta$. In the range of $0.001 \leq \beta \leq 0.1$, with $L \sim L_{A0} \sim 12$ μm obtained from the simulations, the model agrees well with the data.

## Section 1: Support for Fig. 2

Figure S1 shows the characteristics of photoemission collected from a NW/MoS$_2$ hybrid with a fully overlapping NW, as discussed in the main text with respect to Fig. 2. Red emission from the NW [Fig. S1(a)] implies that there are absorption and emission losses from SPP propagation in the MoS$_2$ emitter. The stronger emission at the NW ends was attributed to SPP scattering [1]. Although PL enhancement is relatively negligible, the behavior of selective A$^0$ exciton enhancement is consistent with that in the case of the partially overlapping NW [Fig. S1(b)].

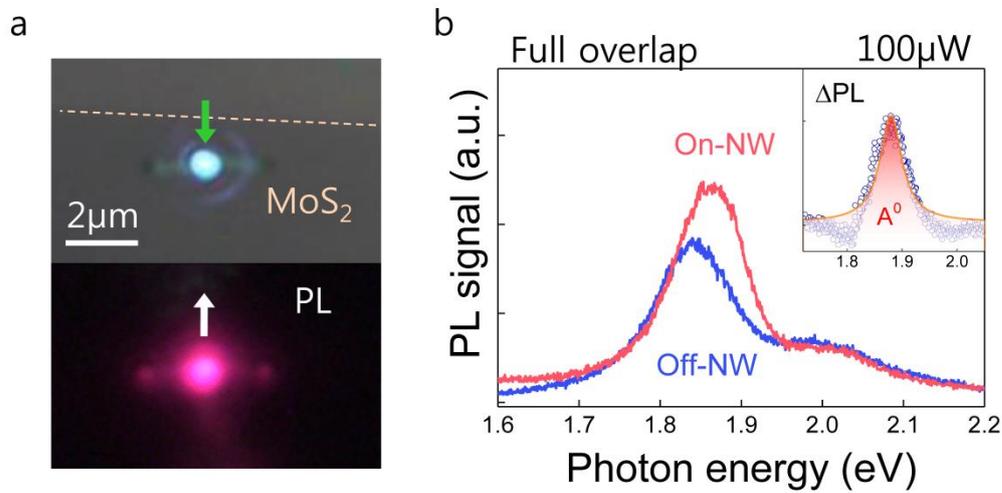

FIG. S1. (a) Optical micrograph and PL image of fully overlapping sample. Green arrow: input laser ($\lambda$ = 514 nm); white arrow: PL signal collection at the same position as the input laser. (b) Comparison of the PL signals of On-NW and Off-NW at 100 μW laser power. Inset: ΔPL (the PL for Off-NW subtracted from that for On-NW), which indicates the enhancement of only the A$^0$ exciton by single Lorentzian fit.

## Section 2: Support for Fig. 3(c)

We attribute the oscillatory fringes below the $A^0$ energy level, as seen in Fig. 3(c) in the main text, to the Fabry-Pérot cavity effect. To see the fringes more clearly, the difference between the PL signal and the Lorentzian fit [Fig. S2(a) and (b)] is Fourier transformed [Fig. S2(c)]. One discrete period in Fig. S2(c) is comparable to the fringe in Fig. S2(b) [38,39]. The results imply the formation of multimodes in the Fabry-Pérot cavity along the long NW. The sum of the frequency-dependent SPP propagation phase $k_{SP}(\omega)L$ and the reflection phase of both sides of the NW, $2\phi_r$, must be $2n\pi$ to obtain a standing wave, where $n$ is an integer [39]. The modes that satisfy the cavity condition survive and display fringes [Fig. S2(d)]; otherwise, the modes are not intensified.

From the cavity fringes, we can deduce the group velocity of the SPPs [38]:

$$v_g = 2L_{NW} c \frac{\Delta\lambda}{\lambda_0^2}$$

where $L_{NW}$ is the NW length (8.6 μm), $c$ is the speed of light, $\lambda_0$ is the wavelength at the center of the interval between two fringes, and $\Delta\lambda$ is interval between two fringes. Using two peaks around 1.75 eV in Fig. S2(b) and (c), the group velocity was calculated to be ~0.4$c$, which agrees with the typical group velocity of 0.5$c$ for an 8-μm-long Ag NW at ~1.75 eV [38].

The quality factor (Q-factor) for the cavity mode was defined as Q = $\lambda/\Delta\lambda$, where $\lambda$ is the wavelength of the center of a fringe peak and $\Delta\lambda$ is the full-width at half-maximum (FWHM). The fringe peak at ~1.84 eV was fitted by the Lorentzian function as shown by the inset in Fig. S2(b); therefore, Q was estimated to be ~27. This low Q-factor is attributed to the poor reflectivity (~0.25) of the SPPs at the ends of the Ag NW [40]. Notably, the cavity effect near A′ (~1.84 eV) indicates that the A′-coupled SPPs cannot be reabsorbed in the $MoS_2$ layer, as discussed in the main text with respect to Figs. 3 and 4.

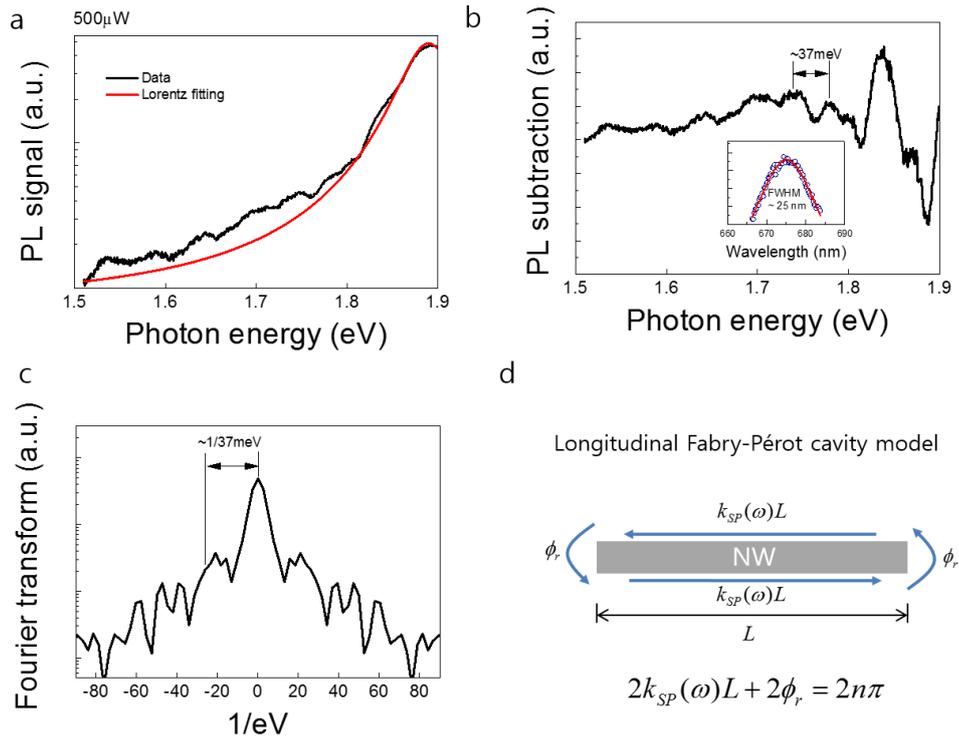

FIG. S2. (a) The PL signal from Fig. 2(c) in the main text and its single Lorentzian fit. (b) The PL signal subtracted from the fitting curve. Inset: Lorentzian fit of the fringe peak as a function of wavelength near 1.84 eV. (c) The Fourier-transformed amplitude curve. (d) Schematic of the longitudinal Fabry-Pérot cavity model of SPPs travelling along the NW.

## Section 3: FDTD numerical simulations

We conducted finite-difference time-domain (FDTD) numerical simulations (Lumerical Solutions, Inc., Vancouver, BC, Canada) for the NW/MoS$_2$ hybrid illustrated in Fig. S3-1(a). The NW/MoS$_2$ hybrid device was on top of a SiO$_2$ (300 nm)/Si substrate with a SiO$_2$ (10 nm) spacer in between. The diameter of the Ag NW was 110 nm, and the MoS$_2$ monolayer was 0.7 nm thick. The minimum mesh size near the MoS$_2$ layer was 0.1 nm. Optical constants $n$ and $k$ for the MoS$_2$ monolayer were taken from [41]. A Gaussian beam (wavelength: 514 nm, beam radius: 180 nm) with polarization modes parallel and perpendicular to the longitudinal NW was used as the input laser source [Fig. S3-1(a)]. The effect of SPP coupling to the NW is negligible when laser input is in the middle of the overlapping NW compared with that for laser input at the distal edge [Fig. S3-1(b)].

This is consistent with the experimental results of a previous report [32]. To activate the waveguide modes of the SPPs ($k_\parallel^{SPP}$) in the NW using laser illumination, $k_\parallel^{SPP} = k_\parallel^0 + \Delta k_\parallel$, where $k_\parallel^0$ denotes the longitudinal component of laser momentum and $\Delta k_\parallel$ is the momentum mismatch in axial direction. To compensate a momentum mismatch, a light-scattering method that provides additional wavevectors can be used. When the laser illuminates the end of the NW, wavevectors are generated in all directions because of strong scattering. $\Delta k_\parallel$ compensates for the momentum mismatch between $k_\parallel^{SPP}$ and $k_\parallel^0$ and thus $k_\parallel^{SPP}$ is activated. However, when the laser illuminates the midsection of the NW, laser scattering for axial direction and thus $\Delta k_\parallel$ are negligible, except for strong laser scattering perpendicular to the NW ($\Delta k_\perp$), because the NW is geometrically symmetrical in the axial direction [32]. Furthermore, the reverse process, radiative emission of SPP modes via scattering, can occur at the NW end but not at the NW midsection. Therefore, PL emission signals are not visible in the midsection of the bare NW, except for the scattering due to defect sites, as shown in the PL images of Figs. 3(a), S4-1, and S4-2(a).

SPP modes were analyzed via simulations performed using the configuration shown in Fig. S3-2(a), where PMMA covered the Ag NW on a SiO$_2$ (310 nm)/Si substrate. SPP propagation length, which is derived from the lowest loss mode among the possible confined modes near the NW [Fig. S3-2(a)], is plotted as a function of photon energy in Fig. S3-2(b). The propagation length decreased rapidly as the photon energy increased.

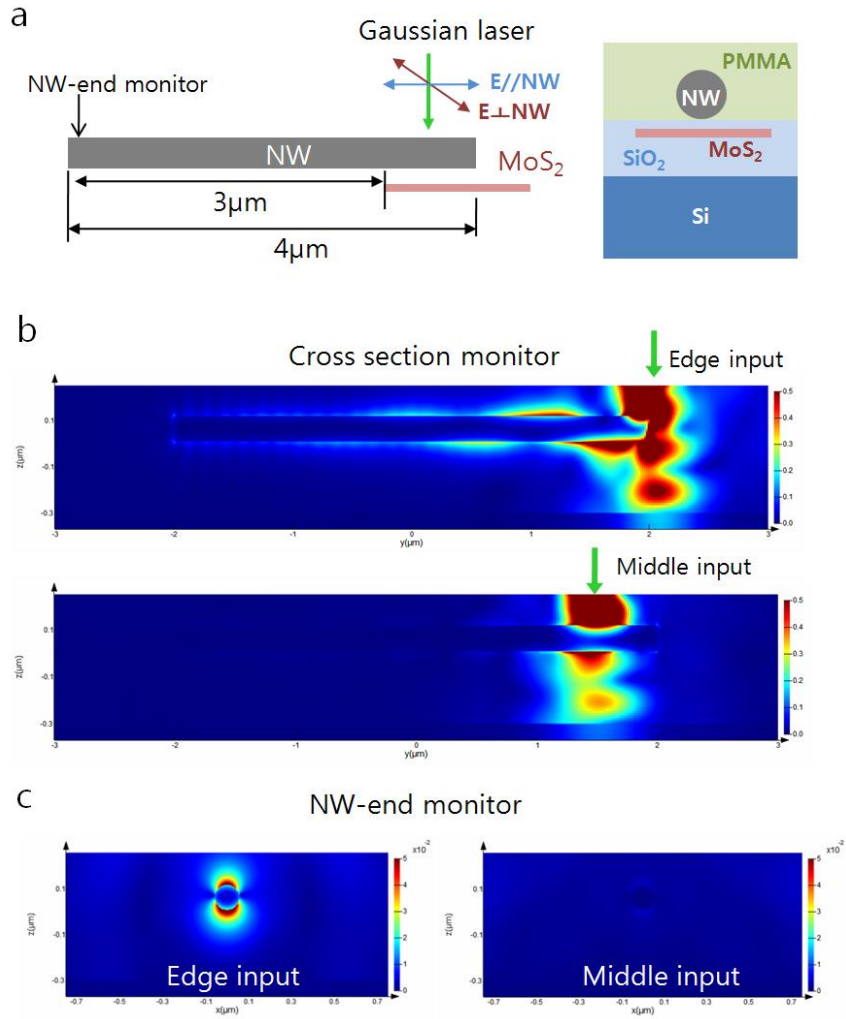

FIG. S3-1. (a) Schematic depicting FDTD simulation conditions and the side view of the NW/MoS$_2$ hybrid. (b) Cross-sectional maps of the electric field intensity along the length of the NW with the laser input at two different positions with respect to the NW. (c) Cross-sectional maps of the electric field intensity as seen looking at the cross section of the NW end.

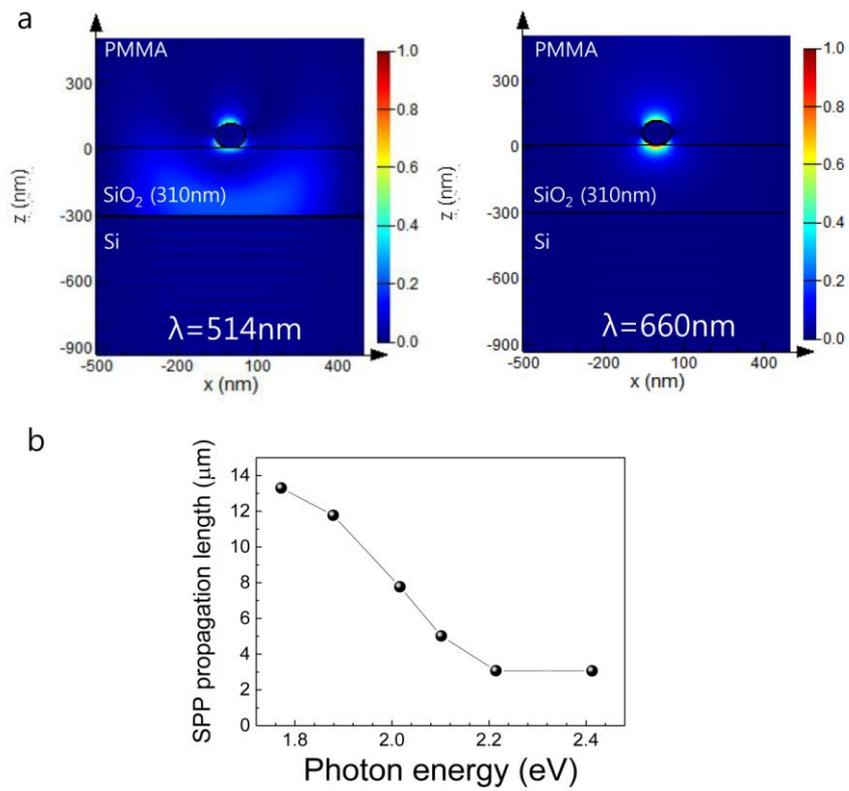

FIG. S3-2. (a) Cross-sectional maps of the electric field intensity at $\lambda$ = 514 and 660 nm. (b) SPP propagation lengths as a function of photon energy were calculated from the imaginary effective index of the SPP mode confined near the NW.

## Section 4: Dependence of laser input position in PL enhancement

To investigate the incident laser coupling effect in $A^0$ exciton enhancement, PL was measured at two different laser positions, i.e., the edge and the middle of the NW, on the partially overlapping NW on MoS$_2$ [Fig. S4-1(a) and (b)] by using the same device as described in the main text for Fig. 3. Figure 3(a) was used for Fig. S4-1(b). The PL images in Fig. S4-1(a) and (b) show that there is strong emission at the laser input position and that the emission decays along the NW that overlaps the MoS$_2$. At the NW-end position [Fig. S4-1(c)], the PL spectrum for the edge input is weaker than that for the middle input. Likewise, at the input position [Fig. S4-1(d)], the PL spectrum shows that the PL enhancement for the middle input is nearly two times greater than that for the edge input. Thus, laser illumination of the middle of the NW causes greater enhancement of the $A^0$ exciton.

According to the FDTD simulation discussed in Section 3 and a previous report [29], the input laser is efficiently coupled to the SPPs when at the NW edge position, but coupling is negligible when the input laser is at the midsection of the NW. This was reconfirmed by the laser position-dependent PL experiments the results of which are shown in Fig. S4-2. The excitons excited by the laser illumination of the NW/MoS$_2$ in the overlapping region are coupled to the NW and propagate along it. Figure S4-2(a) shows that there is strong PL emission at the end of the NW (with extra PL emission at the defect sites in the middle of NW). However, the laser illumination of the midsection of the NW cannot activate the SPPs in the axial direction, as discussed in Section 3. Therefore, no PL emission by the laser-coupled SPPs in the MoS$_2$ layer is observed in Fig. S4-2(b). Conversely, when the laser input is at the NW end, the laser-coupled SPPs can excite the MoS$_2$ excitons, which was demonstrated in [22].

The laser-coupled SPPs at the NW edge are strongly reabsorbed in the MoS$_2$ layer during propagation. Therefore, PL enhancement in our NW/MoS$_2$ hybrids via SPP-exciton coupling contrasts with that via the antenna effect that arises from direct coupling with the input laser [1,8,42]. Moreover, at high laser power, the $A^0$ dominance in our hybrids compared with the A' dominance in bare MoS$_2$ is distinguished by the redshifting of the A peak without PL intensity enhancement due to the phase transition of MoS$_2$ that results from the plasmon-induced hot electron effect of the Au nanoantenna [43].

Figure S4-3 shows the PL spectra for On-NW using various laser input positions in a single hybrid, where $x_{M2}$ is the length obtained by subtracting the bare NW length ($x$) from the effective NW length between the NW end (left) and the laser input position (green arrow). For On-NW, as $x_{M2}$ increases, the $A^0$ is exclusively enhanced and becomes weaker. The top inset of Fig. S4-3 shows that the $\varepsilon$-factor decays exponentially as a function of $x + x_{M2}$, with decay length of ~ 2.3 μm. The SPPs for longer $x + x_{M2}$ more dominantly lose their intensity than for shorter $x + x_{M2}$ due to enhanced Ohmic and MoS$_2$ absorption loss. The $x + x_{M2}$ dependence also implies that the $A^0$ enhancement is associated with longitudinal SPP propagation along the NWs. Furthermore, the decay behavior of $A^0$ enhancement is well contrasted with the result for partially overlapping NW in Fig. 2 (b) which shows a large fluctuation of A peak enhancement due to the varying sample conditions arising from different samples. However, the data of Figure S4-3 does not display any fluctuation behaviour arising from varying sample conditions, because the single NW has same sample conditions at different positions.

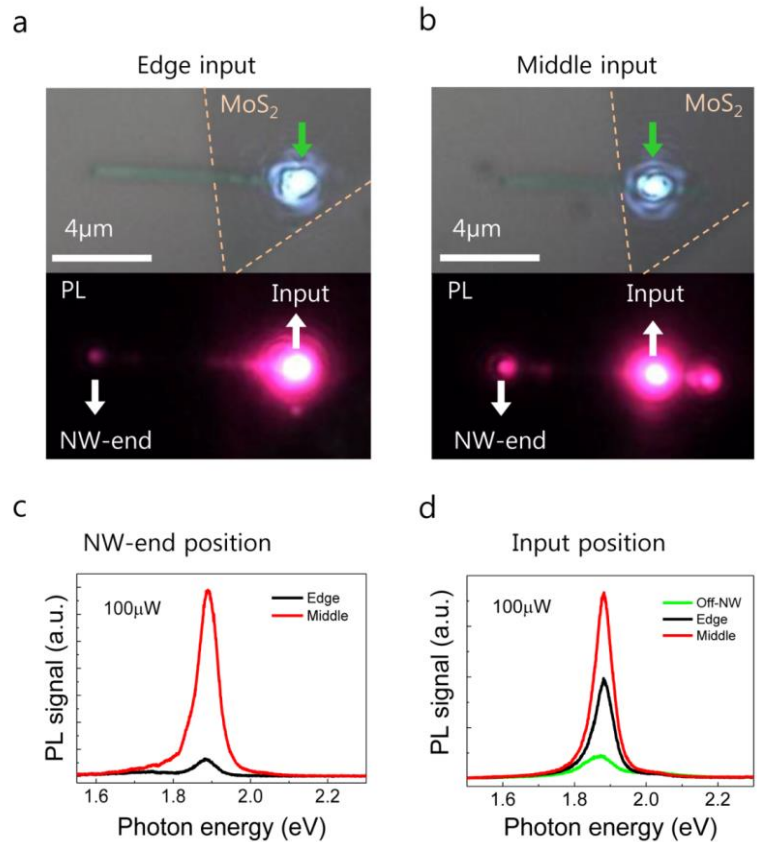

FIG. S4-1. Optical micrographs illustrating the MoS$_2$ flake position and PL images for laser

input positions at (a) the NW edge and (b) the middle of the NW. (c) PL spectra collected at the NW end for the laser input at the NW edge and middle. (d) PL spectra collected at the laser input positions. Green arrow: input laser; white arrow: position of PL signal collection.

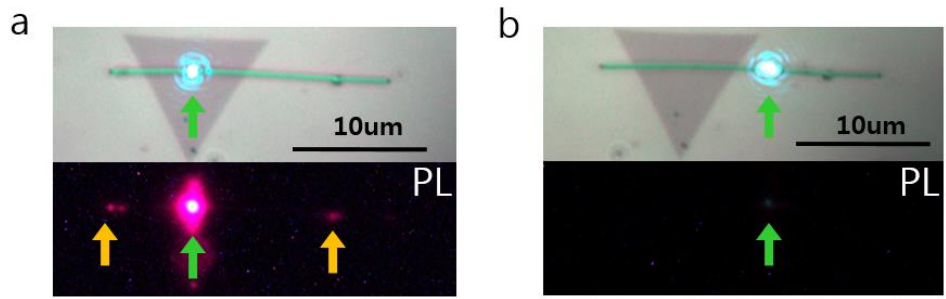

FIG. S4-2. Optical micrographs and PL images for laser input (a) in the NW/MoS$_2$ overlap region and (b) in the middle of the NW but no overlap. Green arrow: input laser; yellow arrow: PL emission via SPP scattering at the NW and at the defect sites in the middle of the NW.

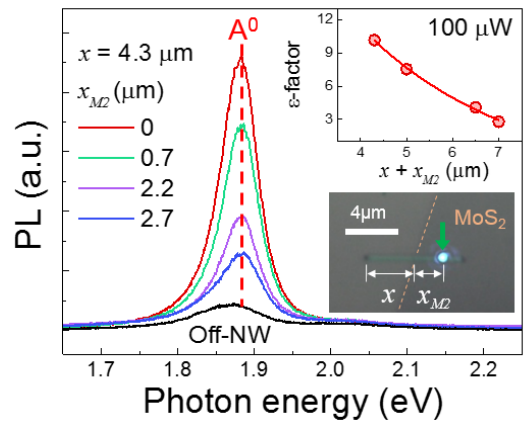

FIG. S4-3 PL spectra with various $x_{M2}$ ($x = 4.3$ μm) in a single hybrid with the length of the NW ~7.5 μm. $x_{M2}$: the subtraction of bare NW length ($x$) from effective NW length between the NW end (left) and the laser input position (green arrow). Top inset graph: $\varepsilon$-factor as a function of $x + x_{M2}$. Bottom inset: optical micrograph showing the definitions of $x$ and $x_{M2}$.

**Section 5: Support of the analytical model for cyclic amplification**

Figure S5(b) shows the PL intensity profile along the NW, extracted from the PL image of the NW/MoS$_2$ hybrid shown in Fig. S5(a). The PL intensity at the NW/MoS$_2$ overlapping side ($x'$) is higher than that of the bare NW side ($x$), because reabsorption and re-emission of SPP in the MoS$_2$ layer occur during propagation. The PL intensity of the left end of the NW is weaker than that of the right end, supporting our assumption that $L \gg L'$ and the amplification effect in the $x'$ area is negligible, as discussed in the main text with respect to Fig. 4.

Figure S5(c) shows each exciton emission situation for Off-NW and for On-NW at the two positions, P1 and P2, in Fig. S5(b). In the Off-NW situation, the band-filling effect generates exciton complexes. In the NW/MoS$_2$ overlapping region, cyclic SPP travelling causes cumulative $A^0$ amplification that results in PL enhancement at the laser input position (P1), while SPPs lose their energy at P2 due to their reabsorption and re-emission during propagation in MoS$_2$. Here, the A'-coupled SPPs cannot be absorbed in the MoS$_2$ layer and generate cavity modes.

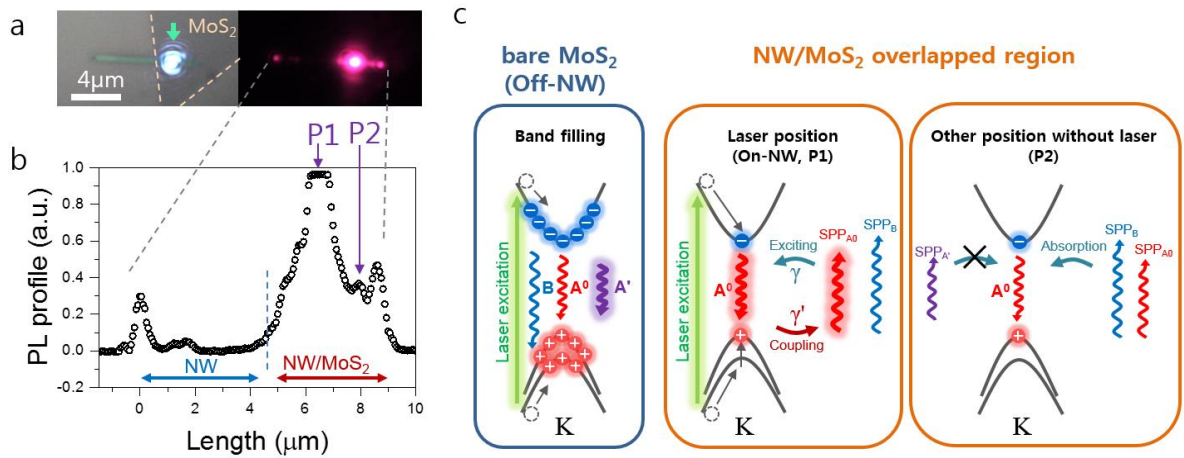

FIG. S5. (a) Optical micrograph and PL image for the NW/MoS$_2$ hybrid. Green arrow: input laser. (b) PL signal profile derived from the PL intensity mapping of the PL image along the NW. P1 is the laser input position and P2 is another position in the NW/MoS$_2$ overlapping region without laser input. (c) Comparison of exciton emission situations for Off-NW and the P1 and P2 positions in (b).

# Supplemental references